\documentclass[preprintnumbers,prd,aps,showpacs,twocolumn,amsmath,amssymb,floats,superscriptaddress,groupedaddress]{revtex4-1}
\usepackage{graphicx}
\usepackage{dcolumn}   
\usepackage{epsfig}
\usepackage{bm}
\usepackage{amsfonts}
\usepackage{amssymb}
\usepackage{times}
\usepackage{natbib}

\usepackage[dvips]{color}

\def\lapp{\mathrel{\rlap{\raise.5ex\hbox{$<$}}
                    {\lower.5ex\hbox{$\sim$}}}}
\def\gapp{\mathrel{\rlap{\raise.5ex\hbox{$>$}}
                    {\lower.5ex\hbox{$\sim$}}}}

\begin{document}
\title{Helical cosmological magnetic fields from extra-dimensions}

\author{Kumar Atmjeet}
\email{katmjeet@physics.du.ac.in}
\affiliation{Department of Physics and Astrophysics, University of Delhi,
Delhi 110007, India.}
\author{T R Seshadri}
\email{trs@physics.du.ac.in}
\affiliation{Department of Physics and Astrophysics, University of Delhi,
Delhi 110007, India.}
\author{Kandaswamy Subramanian}
\email{kandu@iucaa.ernet.in}
\affiliation{IUCAA, Post Bag 4, Ganeshkhind, Pune 411007, India.}
\date{\today}


\begin{abstract}
We study the inflationary generation of helical cosmological magnetic fields in 
a higher-dimensional 
generalization of the electromagnetic theory. For this purpose, we also include a parity breaking piece to the electromagnetic action.
The evolution of extra-dimensional scale factor allows the breaking of conformal invariance of 
the effective electromagnetic action in $1+3$ dimensions required for such generation.  
Analytical solutions for the vector potential  can be obtained in terms of Coulomb wave-functions for some special cases. 
We also present numerical solutions for the vector potential evolution in more general cases.
In the presence of a higher-dimensional cosmological constant there exist solutions for the scale factors in which 
both normal and extra dimensional space either inflate or deflate simultaneously with the same rate. In such a scenario, with 
the number of extra dimensions $D=4$, a scale invariant spectrum of helical magnetic field is obtained. 
The net helicity arises, as one helical mode comes to dominate over the other at the superhorizon scales.
A magnetic field strength of the order 
of $10^{-9}$ $G$ can be obtained for the inflationary scale $H\simeq 10^{-3}$ $M_{pl}$. Weaker fields will be generated for lower scales of inflation. 
Magnetic fields 
generated in this model respects the bounds on magnetic fields by Planck and $\gamma$-ray observations (i.e. $10^{-16}$ $G$ $<$ $B_{obs}<
3.4\times 10^{-9}$ $G$).
\end{abstract}



\maketitle
\section{INTRODUCTION}
A fully satisfactory theory that can explain the origin of cosmic magnetic fields is still elusive. On the observational front, 
we find evidence for magnetic fields over a range of scales including cosmological scales. 
Coherent magnetic fields of the strength of about few $\mu$G are observed at the scales of Kpc to $10$ Kpc in the nearby disk galaxies and galaxy clusters
\cite{Beck12,BS05,ckb01,gof04}. Such fields have also been inferred in galaxies at high redshifts of $z \simeq 1-2$  \cite{kpp08,Bernet08}. 
In the intergalactic medium (IGM), on mega-parsec (Mpc) scales, there are indications of a volume filling field of more 
than $3 \times 10^{-16}$ $G$ \cite{NV10,tlr11}. Several scenarios have been suggested to 
explain the origin of these fields over different scales. For magnetic fields in collapsed structures like galaxies, astrophysical processes could 
suffice in providing an explanation \cite{BS05,Beck12}. On the other hand, a primordial origin appears the most natural and simplest scenario for very 
large scale magnetic fields, especially one that volume fills the IGM.

Primordial magnetic fields with large coherence scales could possibly originate in the early universe \cite{kpp93,gras01,dol03,widrow_etal12,durn13}.
One promising route to understand the origin of these fields is via 
a mechanism in the inflationary context \cite{turn88,rat92}. However, to generate large enough fields during inflation, one also needs 
to break the conformal invariance of the electromagnetic action. A simple way by which this has been implemented is to introduce a 
coupling of the electromagnetic Lagrangian density to a scalar function of some dynamical variables like the inflaton, curvature etc 
\cite{wid02,gio2000,gio04,gio08,gios08,dol93,kkt11,kunze05,turn88,rat92,dur06,cdf09,ll95,ggv95,bamba04,bamba07,mukh09,bptv01,ccf08,my08,kandu10}.
However, in many models there is no fundamental reason to introduce this coupling factor other than the fact that the conformal invariance of 
the electromagnetic action needs to be broken. 

A natural way to break conformal invariance of the electromagnetic action, 
at least in the early universe, can be envisaged in a cosmology motivated by higher-dimensional theories \cite{gio2000,kkt11,kunze05,gios08}. 
In this approach, one starts with a higher-dimensional cosmology (in $1+3+D$-dimensional space-time). The action contains higher-dimensional 
generalization of the electromagnetic gauge field. On dimensional reduction, parameters of the higher dimensions 
(more specifically the scale factor of the higher-dimensional space), naturally appear as multiplicative factors to the $1+3$-dimensional 
electromagnetic Lagrangian. Since the parameters of the higher dimension evolves with time, the breaking of conformal invariance is ensured.

Moreover, it was shown in \cite{akis14}, that a natural way to have both these features (namely, breaking of conformal 
invariance of the electromagnetic action as well as an inflationary phase in the early universe) is by considering higher-dimensional 
action with a Gauss-Bonnet term. In this approach the postulate for a scalar field is neither required for breaking conformal invariance 
nor for realizing inflation. 
In this work we have expanded upon the earlier work \cite{akis14}, by adding a helical term to the action in the context of higher-dimensional 
theories. This allows for the potential generation of helical cosmological magnetic fields. Note that in the context of $1+3$-dimensional 
theories possible generation of helical magnetic fields during inflation is discussed by several authors \cite{cdf09,rjdh11,rjbh12,rjdh12,cap14,cheng14}. 

Large scale primordial helical magnetic fields are also interesting from another
aspect. Note that primordial fields captured into collapsed objects will be
subjected to turbulent diffusion. This can lead to a rapid dissipation of the field
if it was non helical \citep{RSS88}. However, it turns out that large scale helical fields
are resilient to such turbulent diffusion, due to magnetic helicity conservation,
and only decay on the slow resistive time scale \citep{KBJ11,kb13,bkb14}. This makes such helical fields
more relevant even in collapsed objects like galaxies and galaxy clusters.

The plan of the paper is as follows. In Sec. \ref{EMT}, we briefly describe the problem of generating electromagnetic fields in 
$1+3$-dimensional theory with standard electromagnetic action. 
We also describe there the formulation of the electromagnetic theory in the higher-dimensional space-time. The background model of space-time is motivated from
Gauss-Bonnet gravity which is discussed in Sec. \ref{GB}.
In Sec. \ref{HF} we introduce a parity breaking term to the electromagnetic action which may lead to the generation of the 
helical magnetic fields.
The detailed analytical and numerical solutions are discussed in the Secs. \ref{ANALYTICAL} and \ref{NUMERICAL}. We estimate the strength of 
the magnetic fields obtained in our model in Sec. \ref{MF} before concluding in Sec. \ref{conc}. 

The notations and conventions used in this work are as follows. We work in natural units (i.e. $\hslash$ $=$ $G$ $=$ $c$ $=$ $1$). 
We chose the metric signature to be ($-,+,+,+,+....$). Lowercase Latin indices run from $1$ to $3$ while the uppercase Latin indices take values from $4$ to 
$3+D$, where $D$ is the number of extra dimensions in our model. The Greek alphabets can take values from $0$ to $1+3+D$.


\section{ELECTROMAGNETIC ACTION IN HIGHER-DIMENSIONAL MODELS}\label{EMT}
The action for the electromagnetic field in a general $1+3$ dimensions is given by,
\begin{equation}
 S^{1+3}_{EM}=-\int{\frac{1}{16 \pi}d^4x \sqrt{-g} F_{\mu \nu}F^{\mu \nu}},
 \label{EM}
\end{equation}
 where $F_{\mu \nu}$ is the electromagnetic field tensor given in terms of the derivatives of vector potential $A_{\mu}$, as 
 $F_{\mu \nu}=\partial_{\mu}A_{\nu}-\partial_{\nu}A_{\mu}$.
 Here, $g$ is the determinant of the metric tensor $g_{\mu \nu}$. At any epoch, the spatial part of this metric is considered 
 to be homogeneous and isotropic. For a homogeneous and isotropic universe, the space-time metric is described by the line element,
\begin{equation}
 ds^2=a(\eta)^{2}(-d\eta^2+\eta_{ij}dx^{i}dx^{j}).
 \label{flat}
\end{equation}
Here $a(\eta)$ is the scale factor of the universe and $\eta_{ij}=diag(1,1,1)$, is the spatial part of Minkowski metric tensor. 
Further, conformal time $\eta$ is related to the comoving time $t$ by,
\begin{equation}
 \eta=\int{\frac{dt}{a(t)}}.
 \label{etaref}
\end{equation}
Since electromagnetic action in Eq.~(\ref{EM}) is conformally invariant, it can be shown in general
that in such a conformally flat background described by the metric in Eq.~(\ref{flat}) electric (E) and magnetic fields (B) will decay as $1/a^{2}$. 
Therefore at the end of inflation such fields will be negligible in strength. In order to have a significant generation of 
electromagnetic fields during inflation, we necessarily need to break this conformal invariance of electromagnetic action. 
Put alternatively, one requires the amplification of $a^2B$. Several such mechanisms to break conformal invariance for magnetogenesis have 
been investigated in literature
\cite{wid02,gio04,gio08,gios08,dol93,kkt11,turn88,rat92,dur06,cdf09,ll95,ggv95,bamba04,bamba07,mukh09,bptv01,ccf08,my08,kandu10}. 
For example, the breaking of conformal invariance can be achieved by introducing a time dependent coupling function prefixing
$F^{\mu \nu}F_{\mu \nu}$, instead of a constant as in the standard electrodynamics.

A natural scenario for breaking conformal invariance and generating primordial magnetic fields arises in the context of higher-dimensional 
theories \cite{kkt11,kunze05,gios08,gio2000}.
We have earlier explored this possibility, where extra-dimensional model with a Gauss-Bonnet term provides a coupling function for breaking of conformal invariance 
in the reduced four dimensional action \cite{akis14}. We now extend this work by  adding a parity breaking piece to the action, which 
allows for the generation of helical primordial magnetic fields. Specifically we consider a higher-dimensional space-time which 
has $D$ extra spatial dimensions in addition to the normal $1+3$ dimensions.
We further assume that the spatial part of normal as well as extra-dimensional subspaces are homogeneous, isotropic and flat.
The line-element for such a universe is be given by,
\begin{eqnarray}
 ds^2 & = & \tilde{g}_{\mu\nu}dx^{\mu}dx^{\nu} \nonumber \\
      & = & -dt^2+a^2(t)\eta_{ij}dx^{i}dx^{j}+b^2(t) \eta_{IJ}dx^{I}dx^{J},
\label{metric}
\end{eqnarray}
where $a(t)$ and $b(t)$ are the scale factors of normal and extra dimensions respectively and $\tilde{g}_{\mu \nu}$ is the higher-dimensional metric. 
We take the action for electromagnetic fields in higher dimensions to be given by,
\begin{equation}
S_{EM}=\frac{-1}{16 \pi}\int{d^{4+D}x\sqrt{-\tilde{g}} \tilde{\mathcal{L}}_{EM}}.
\label{emaction}
\end{equation}
Here,  $\tilde{\mathcal{L}}_{EM}$ is the Lagrangian 
density of the electromagnetic field in higher dimensions and is given by,
\begin{equation}
\bar{\mathcal{L}}_{EM} = \ell \tilde{F}_{\mu \nu}\tilde{F}^{\mu \nu}-\bar{f}\tilde{F}_{\alpha \beta}
 \tilde{F}^{* \alpha \beta}. 
 \label{lag}
\end{equation}
We have introduced two arbitrary time dependent scalar functions, $\ell$ and $\bar{f}$ to keep the action quite general. Further, $\tilde{g}$ 
is the determinant of higher-dimensional metric, $\tilde{g}_{\mu \nu}$. 
The higher-dimensional electromagnetic field tensor is expressed in terms of higher-dimensional vector potentials, $\tilde{A}_{\mu}$. We 
have also defined the dual of the higher-dimensional electromagnetic tensor $\tilde{F}^{* \alpha \beta}$ as,
\begin{equation}
 \tilde{F}^{* \alpha \beta}= \tilde{\eta}^{\alpha \beta \gamma \delta \psi \phi....4+D} \tilde{F}_{\gamma \delta}{Q}_{\psi \phi.....4+D}.
 \label{levi}
\end{equation}
Here, $\tilde{\eta}^{\alpha \beta \gamma \delta \psi \phi....4+D}$ is the higher-dimensional Levi-Civita tensor and ${Q}_{\psi \phi.....4+D}$
is the tensorial field needed to describe the $4+D$ dimensional dual that we have defined. Note that the term $\bar{f}\tilde{F}_{\alpha \beta}
 \tilde{F}^{* \alpha \beta}$ in Eq.~(\ref{lag}) is a parity breaking term which could lead to the generation of helical magnetic fields.


\section{Gauss-Bonnet Gravity}\label{GB}
We assume that the dynamics of the universe is governed by the action of the form  \cite{abm07,chto12,pahwa08,akis14},
\begin{eqnarray}
 S \, & = & \, \int d^{4+D}x \sqrt{-\tilde{g}} \,  \left(\mathcal{L}_{matter}-\frac{1}{16\pi}\tilde{\mathcal{L}}_{EM}\right. \nonumber \\
 \, & &  \, \left. -\frac{M^{D+2}}{2}(\tilde{R}\,+ \chi \, \tilde{G})+\bar{\Lambda} \right)
 \label{action}.
\end{eqnarray}
Here, $M$ is the higher-dimensional Planck mass which is related to $1+3$-dimensional Planck mass ($M_{pl}$) as, $M^{D+2}b^{D}=M_{pl}^{2}$. 
$\tilde{R}$ is the $1+3+D$-dimensional Ricci scalar and $\chi$ is the Gauss-Bonnet parameter for the Gauss-Bonnet term ($\tilde{G}$) given by,
\begin{equation}
 \tilde{G}=\tilde{R}^2-4\tilde{R}_{\mu \nu}\tilde{R}^{\mu \nu}+\tilde{R}_{\mu \nu \lambda \sigma}\tilde{R}^{\mu \nu \lambda \sigma}.
\end{equation}
A cosmological constant term ($\bar{\Lambda}$) has been added in the above action in order to keep it general, and also because it leads to some interesting 
cosmological models \cite{pahwa08,akis14}.
While in $1+3$-dimensional gravity the Gauss-Bonnet term becomes a total divergence (and hence doesn't contribute to the equation of motion), in
higher dimensions, it gives a nonzero contribution . Further, as the Gauss-Bonnet term varies as square of the curvature, it has no 
significant contribution on cosmological scales at the present epoch.

Solutions for scale factors $a(t)$ and $b(t)$ in this scenario have been discussed in \cite{pahwa08,akis14}. The asymptotic behavior of the scale factors are 
exponential in time. In addition to the solutions in which normal dimension inflates and extra dimension deflates or vice-versa, the inclusion 
of  a cosmological constant $\bar{\Lambda}$ gives interesting solutions in which both the scale factors ( i.e. of normal and extra dimensions) either 
increase or decrease, simultaneously. The solutions in general are given by,
\begin{equation}
 a(t) \propto  e^{\alpha t}, \qquad b(t) \propto e^{\beta t}.
 \label{scales}
\end{equation}
Here $\alpha$ and $\beta$ are the exponents for scale factors $a(t)$ and $b(t)$ respectively. The signs of these exponents determine which of the 
spatial dimensions (normal or extra)
are inflating or contracting. A detailed set of acceptable solutions in this scenario are discussed in \cite{akis14,pahwa08}.


\section{HELICAL MAGNETIC FIELDS IN HIGHER-DIMENSIONAL COSMOLOGY}\label{HF}
We have considered the metric given in Eq.~(\ref{metric}) to describe the extra-dimensional universe. We impose gauge conditions on higher-dimensional 
vector potential by adopting,
$\tilde{A}_I=0$ and ${\partial}_{I}\tilde{A}_{\mu}=0$ \cite{gio2000}. This choice of gauge ensures that only $1+3$-dimensional components of the vector 
potential $\tilde{A}^{\mu}$ are nonzero and further they depend only on the co-ordinates of normal dimensions. With these 
gauge conditions a dimensional reduction of the electromagnetic part of the action in Eq.~(\ref{action}), gives an 
effective $1+3$-dimensional electromagnetic action, 
\begin{equation}
 S_{em}=-\int{\frac{1}{16\pi}d^{4}x \sqrt{-g}\left(\frac{b}{b_{0}}\right)^{D}\mathcal{L}_{EM}}.
 \label{redaction}
\end{equation}
Here, $\mathcal{L}_{EM}=b_{0}^{D}\Omega_{D}\tilde{\mathcal{L}}_{EM}$.
$\Omega_{D}$ is the co-ordinate volume of extra dimensions which is assumed to be finite and $g$ is the determinant of
$1+3$ -dimensional metric tensor $g_{\mu\nu}$. From the definition of $\tilde{{F}^{*}}^{\mu \nu}$ in Eq.~(\ref{levi}) and the gauge conditions
on vector potential, one sees that indices $\psi$,$\phi$, etc can take values purely of the extra-dimensional space. We may note that $Q_{\psi \phi...4+D}$
is purely an antisymmetric tensor. These two conditions imply that there is only one independent component of $Q_{\psi \phi...4+D}$ that comes into 
the reduced action. We combine this with $\bar{f}$ to define a new function, $f$ . We can then write,
\begin{displaymath}
 \tilde{\eta}^{\alpha \beta \gamma \delta \psi ...4+D}Q_{\psi \phi...4+D}= {\eta}^{\alpha \beta \gamma \delta},
\end{displaymath}
where ${\eta}^{\alpha \beta \gamma \delta}$ is the usual $1+3$-dimensional Levi-Civita tensor.
Therefore, $\mathcal{L}_{EM}$ is the equivalent $1+3$ -dimensional Lagrangian density for $1+3$-dimensional vector potential $A_{\mu}$ ($\mu= 0$ to 
$3$ henceforth) defined by,
\begin{equation}
 \mathcal{L}_{EM}=\left[\ell F_{\mu\nu}F^{\mu\nu}-f F_{\alpha \beta}
 F^{* \alpha \beta}\right].
\end{equation}
The $1+3$- dimensional vector potential corresponding to this $1+3$-dimensional electromagnetic action
is given by, 
\begin{equation}
 A_{i}=(\Omega_{D}b_{0}^{D})^{1/2}\tilde{A}_{i}.
\end{equation}
The reduced $1+3$-dimensional electromagnetic action is no longer conformally invariant because of the time dependent function $(b/b_{0})^D$ in 
Eq.~(\ref{redaction}) coupling to 
 $\mathcal{L}_{EM}$ (even if $\ell$ and $f$ are constants). By varying the $1+3$-dimensional electromagnetic action with respect to $1+3$-dimensional 
vector potential we obtain 
Maxwell's equations as,
\begin{equation}
 \partial_{\mu}\left[\sqrt{-g}\left(\frac{b}{b_{0}}\right)^{D}\ell{F}^{\mu\nu}\right]=\partial_{\alpha}\left[\sqrt{-g}\left(\frac{b}{b_{0}}\right)^{D}f
 \epsilon^{\alpha \beta \gamma \delta}{F}_{ \gamma \delta}\right]
 \label{maxwell}
\end{equation}
We chose to work in radiation gauge i.e. $A_{0}=0$, $\partial_{i}A^{i}=0$. We also use the fact that, Maxwell's equations for the dual is the 
identity, $\partial_{\alpha}\left[\epsilon^{\alpha \beta \gamma \delta}{F}_{ \gamma \delta}\right]=0$. Moreover, it is convenient to work in terms 
of conformal time co-ordinate $\eta$. Maxwell's equation, Eq.~(\ref{maxwell}) then takes the form,
\begin{eqnarray}
 A_{j}''(\eta,x) & + &\left[D\frac{b'}{b}+\frac{{\ell}'}{\ell}\right]A_{j}'(\eta,x)-\partial_{i}\partial_{i}A_{j}(\eta,x) \nonumber \\
 & - & \left[D\frac{b'}{b}\frac{f}{\ell}  + \frac{f'}{\ell}\right]\frac{1}{2}\epsilon^{0j\phi\psi}F_{\phi\psi}=0
\end{eqnarray}
where prime is the derivative with respect to $\eta$. 
It can be seen that the presence of dynamical extra-dimensional scale factor as well as time-dependent functions $\ell$ and $f$ break
the conformal invariance of electromagnetic action in $1+3$-dimensions which may amplify the electromagnetic field fluctuations. The formalism 
till now is for general $\ell$ and $f$. In this work we will explore purely the effects of extra dimensions, 
i.e.  we chose that $\ell=f=1$. For this particular case the equation for the vector potential reduces to,
\begin{equation}
 A_{j}''+D\frac{b'}{b}A_{j}-\partial_{i}\partial_{i}A_{j}-\frac{D}{2}\frac{b'}{b}\epsilon^{0j\psi\phi}F_{\psi\phi}=0
\end{equation}
In order to quantize the vector potential we express it in terms of its Fourier components $A_{h}(k,\eta)$ as (cf. \cite{rjdh11,cdf09}), 
\begin{eqnarray}
 A_{l}(x,t)& = &\sqrt{4 \pi}\int \frac{d^3k}{(2\pi)^3}\sum_{h,\lambda=1}^2 {\bf \epsilon}_{h l}^{\bf k}\left[b_{\lambda}({\bf k})A_{h}(k,\eta) e^{i{\bf k}\cdot {\bf x}}\right.  \nonumber \\
 & + & \left. b_{\lambda}^{\dag}(k)A_{h}^{*}(k,\eta)e^{-i{\bf k}\cdot {\bf x}}\right].
\end{eqnarray}
Here, we have defined the helicity basis as,
\begin{equation}
 {\bf \epsilon}_{h}^{{\bf k}}=\frac{1}{\sqrt{2}}\left({\bf \epsilon}_{1}^{\bf k}+hi{\bf \epsilon}_{2}^{\bf k}\right),
\end{equation}
with $h = \pm 1$ denoting positive and negative helicities. Also ${\bf \epsilon}_{1}^{\bf k}$ and ${\bf \epsilon}_{2}^{\bf k}$ are the two transverse polarization vectors corresponding to corresponding to 
$\lambda=1$ and $2$ respectively. These two polarization vectors along with $\hat{{\bf k}}$ form the orthonormal spatial basis as,
\begin{equation}
 \left({\bf \epsilon}_{1}^{\bf k}, {\bf \epsilon}_{2}^{\bf k}, \hat{{\bf k}}\right),\hspace{1cm} |{\bf \epsilon}_{\lambda}^{\bf k}|^{2}=1,\hspace{1cm} \hat{{\bf k}}=\frac{{\bf k}}{k}
\end{equation}
Further, $b_{\lambda}(k)$ and $b_{\lambda}^{\dag}(k)$ are the annihilation and creation operators, which satisfy the commutation relations,
\begin{eqnarray}
 & &[b_{\lambda}(k),b_{\lambda'}^{\dag}(k)]=\delta_{\lambda,\lambda'}\delta^3(k-k'), \nonumber \\
 & &[b_{\lambda}(k),b_{\lambda'}(k)]=[b_{\lambda}^{\dag}(k),b_{\lambda'}^{\dag}(k)]=0
\end{eqnarray}
The Fourier co-efficients for helicity modes, $\bar{A}_{h}(k,\eta)$ (defined as $a A_{h}(k,\eta)$) satisfy the equation
\begin{eqnarray}
\bar{A_{h}}''(k,\eta)&+&D\frac{b'}{b}\bar{A_{h}}'(k,\eta)+k^2\bar{A_{h}}(k,\eta) \nonumber \\
 &-&D\frac{b'}{b}hkA_{h}(k,\eta)=0. 
\label{heleq}
\end{eqnarray}
Here $h$ denotes helicity of the modes depending on signs. Defining a new variable $q(\eta)$ such that, $D\frac{b'}{b}=2\frac{q'(\eta)}{q(\eta)}$. 
We can rewrite Eq.(\ref{heleq}) in terms of $\mathcal{A}_{h}(k,\eta)$ as,
\begin{equation}
  \mathcal{A}_{h}''(k,\eta)+\left[k^2-\frac{q''(\eta)}{q(\eta)}-2\frac{q'(\eta)}{q(\eta)}hk\right]\mathcal{A}_{h}(k,\eta)=0,
  \label{VP}
\end{equation}
where, $\mathcal{A}_{h}(k,\eta)=q(\eta)\bar{A}_{h}(k,\eta)$. 
As the solution for the scale factors are exponential given by, Eq.~(\ref{scales}), we have
\begin{equation}
 q(\eta)\propto \eta^{-\frac{\beta D}{2 \alpha}}.
\end{equation}
Therefore, Eq.(\ref{VP}) takes the form,
\begin{equation}
 \mathcal{A}_{h}''(k,\eta)+\left[k^2-V(\eta)-V_{h}(\eta)\right]\mathcal{A}_{h}(k,\eta)=0,
 \label{Ah}
\end{equation}
where,
\begin{eqnarray}
 V(\eta) & = & \left(\frac{\beta D}{2 \alpha}\right)\left(\frac{\beta D}{2 \alpha}+1\right)\frac{1}{\eta^{2}},\nonumber \\
 V_{h}(\eta)& = & -2\left(\frac{\beta D}{2 \alpha}\right)\frac{hk}{\eta}
\end{eqnarray}


\section{ANALYTICAL SOLUTIONS FOR VECTOR POTENTIAL} \label{ANALYTICAL}
Defining a new variable $z=-k\eta$ we can transform Eq.~(\ref{Ah}) into,
\begin{equation}
 \frac{d^2\mathcal{A}_{hi}}{dz^{2}}(z)+\left[1-\frac{m_{i}(m_{i}+1)}{z^{2}}-\frac{2nh}{z}\right]\mathcal{A}_{hi}(z)=0.
 \label{coul}
\end{equation}
Here, $i=1$ or $2$, $m_{1}=\frac{\beta D}{2 \alpha}$ (for $\frac{\beta D}{2 \alpha}>0$) and $m_{2}=-\frac{\beta D}{2 \alpha}-1$ (for
$\frac{\beta D}{2 \alpha}<-1$). Further, we define $\beta D/2\alpha=n$. Eq.~(\ref{coul}) represents Coulomb's equation whenever $m_{i}$ is a positive integer.
For scales much smaller than the horizon size, i.e. $-k\eta>>1$, we would like to match it with  the outgoing wave solution in the Bunch-Davis vacuum given by,
\begin{equation}
 \mathcal{A}(k,\eta)=\frac{1}{\sqrt{2k}}\exp{-ik\eta}.
 \label{pln}
\end{equation}
The generic solutions for Eq.~(\ref{coul}) are given as the linear combinations of regular and irregular Coulomb wave-functions, $F_{m_{i}}(hn,z)$ and 
$G_{m_{i}}(hn,z)$, \cite{1965stegunA}. For $z\rightarrow  \infty$ i.e. the sub horizon regime, we have,
\begin{equation}
 G_{m_{i}}(hn,z)\pm iF_{m_{i}}(hn,z)\sim e^{\pm iz}.
\end{equation}
The solution for the vector potential with the required sign of the outgoing wave function is,
\begin{equation}
 \mathcal{A}_{hi}=\frac{1}{\sqrt{2k}}\left[G_{m_{i}}(hn,z)+ iF_{m_{i}}(hn,z)\right]
 \label{sol}
\end{equation}
For modes which go outside the horizon, we have $z\rightarrow 0$, and in this case, the asymptotic behavior of Coulomb functions is 
given by \cite{1965stegunA},
\begin{eqnarray}
 F_{m_{i}}(hn,z) & \rightarrow & 0,\\
 G_{m_{i}}(hn,z) & \rightarrow & \frac{2(2nh)^{m_{i}}}{C_{m_{i}}(nh)(2m+1)!}(2nhz)^{1/2}\nonumber \\
 & & \times K_{2m_{i}+1}[2(2nhz)^{1/2}]
\end{eqnarray}
with, 
\begin{equation}
 C_{m_{i}}(nh)=\frac{2^{m_{i}} e^{-\frac{\pi h n}{2}}\left|\Gamma(m_{i}+1+ihn)\right|}{\Gamma(2m_{i}+2)}.
 \label{cm}
\end{equation}
Here, $K_{2m_{i}+1}(2(2nhz)^{1/2})$ is the modified Bessel's function whose asymptotic form for $ z\rightarrow 0$ is given by,
\begin{equation}
 K_{2m_{i}+1}(2(2nhz)^{1/2}) \sim \frac{1}{2} (2nhz)^{\frac{-(2m_{i}+1)}{2}}\Gamma(2m_{i}+1).
 \label{bessel}
\end{equation}
Therefore, the solution obtained for super horizon modes can be written as,
\begin{equation}
 \mathcal{A}_{hi}(k,\eta)=\frac{1}{\sqrt{2k}}\frac{\Gamma(2m_{i}+1)}{(2m_{i}+1)!C_{m_{i}}(nh)}(-k\eta)^{-m_{i}}.
\label{Ahi}
 \end{equation}

We can now compute the power spectrum for the generated magnetic fields. This is 
given by \cite{rjdh11,rjbh12,rjdh12,cap14},
\begin{equation}
 \frac{d\rho_{B}}{dlnk}=\frac{1}{(2 \pi)^2}\left(\frac{b}{b_{0}}\right)^{D}k\frac{k^4}{a^2}P_{s}(k,\eta)dk,
 \label{PB}
\end{equation}
where, 
\begin{equation}
 P_{s}(k,\eta)=\left|A_{+}(k,\eta)\right|^2+\left|A_{-}(k,\eta)\right|^2
\end{equation}
Similarly helicity is measured by the antisymmetric combination of power in the different helicity modes. i.e.
\begin{equation}
 \frac{d\rho_{h}}{dlnk}=\frac{1}{(2 \pi)^2}\left(\frac{b}{b_{0}}\right)^{D}k\frac{k^4}{a^2}P_{a}(k,\eta)dk,
 \label{PH}
\end{equation}
where, 
 \begin{equation}
 P_{a}(k,\eta)=\left|A_{+}(k,\eta)\right|^2-\left|A_{-}(k,\eta)\right|^2.
 \end{equation}
In terms of $\mathcal{A}_{h}(k,\eta)$, Eq.~(\ref{PB}) \& ~(\ref{PH}) becomes,
\begin{equation}
\frac{d\rho_{B}}{dlnk} = \frac{1}{(2 \pi)^2}k\left(\frac{k}{a}\right)^{4}\left[\left|\mathcal{A}_{+}(k,\eta)\right|^2+\left|\mathcal{A}_{-}(k\eta)\right|^2\right],
\label{rhob}
\end{equation}
\begin{equation}
\frac{d\rho_{h}}{dlnk} = \frac{1}{(2 \pi)^2}k\left(\frac{k}{a}\right)^{4}\left[\left|\mathcal{A}_{+}(k,\eta)\right|^2-\left|\mathcal{A}_{-}(k,\eta)\right|^2\right].
\label{rhoh}
 \end{equation}

For exponential inflation $k/aH=-k\eta$, where $H$, the Hubble parameter remains constant. We rewrite 
the expression for power spectrum given in Eq.~(\ref{rhob}) as,
\begin{equation}
 \frac{d\rho_{B}}{dlnk} = \frac{k}{(2 \pi)^2}H^{4} (-k\eta)^{4}\left[\left|\mathcal{A}_{+}(k,\eta)\right|^2+\left|\mathcal{A}_{-}(k,\eta)\right|^2\right],
\end{equation}
On super horizon scales, with $m_{i}$ being a positive integer, ratio of the power spectrum in positive helicity mode to that of modes with negative helicity turns 
out to be $e^{2 n\pi}$. Therefore, for $\beta D/2\alpha>0$ positive helicity modes dominate whereas, for $\beta D/2\alpha<0$ the modes with negative
helicity dominate. This is also shown in the numerical solutions obtained in the next section. Substituting Eq.~(\ref{Ahi}) in Eq.~(\ref{rhob})
and neglecting the contribution from the subdominant helicity mode, the power spectrum is then given by,
\begin{equation}
 \frac{d\rho_{B}}{dlnk} = \frac{1}{(2 \pi)^2}H^{4} (-k\eta)^{n_{B}}\left|\frac{1}{\sqrt{2}}\frac{\Gamma(2m_{i}+1)}{(2m_{i}+1)!C_{m_{i}}(nh)} \right|^2.
\label{spectrum}
 \end{equation}
Here the spectral index $n_{B}$ is given by,
\begin{equation}
 n_{B}=4-2m_{i}.
\end{equation}
We see, from Eq.~(\ref{spectrum}) that for $m_{i}=2$, the spectral index, $n_{B}=0$ and hence, this choice leads to a perfect scale invariant power
spectrum for magnetic fields. We get $m_{i}=2$ for $\beta D/2\alpha=2$ or 
$\beta D/2\alpha=-3$. Moreover, this scale invariant spectrum is obtained now for an almost fully helical field.
Note that for arbitrary $\alpha$ and $\beta$, $m_{i}$ may not be a positive integer. Also the integral values of $m_{i}$ limits the 
choice of coupling functions. Thus more general cases are also considered in the next section by numerically solving Eq.~(\ref{Ah}).


\section{NUMERICAL SOLUTION} \label{NUMERICAL}
We first focus on the cases where $m_{i}$ is indeed an integer, as in this case analytical results are available to check the numerics. 
This includes importantly the case when $\alpha=\beta$, $D=4$ (i.e. $\beta D/2\alpha=2$), which as we saw leads to a scale invariant spectrum for the 
magnetic field. Because of the exponential inflation 
of scale factor we take \cite{akis14},
\begin{equation}
 a(\eta)=a_{0}\left(\frac{\eta_{0}}{\eta}\right).
\end{equation}
For numerical calculations we assume $a_{0}=1$ and $\eta_{0}=-1$ without any loss of generality. The ratios of scales 
$H^{-1}$(Hubble length scale) and $a/k$ (length scales for modes) is given by, $k/aH=-k\eta$.
We have written a Mathematica code to obtain the numerical solution of Eq.~(\ref{Ah}). Initial conditions are set for modes well within 
the horizon (i.e. $-k\eta=10$). Solutions are assumed to be plane waves in this region as in Eq.~(\ref{pln}). The  solutions are then obtained at the epoch 
when these modes are much larger than the horizon (i.e. $-k\eta=0.01$). 

\begin{figure}[!htbp]
\centering
{
\includegraphics[width=3.1in,height=2.1in,angle=0]{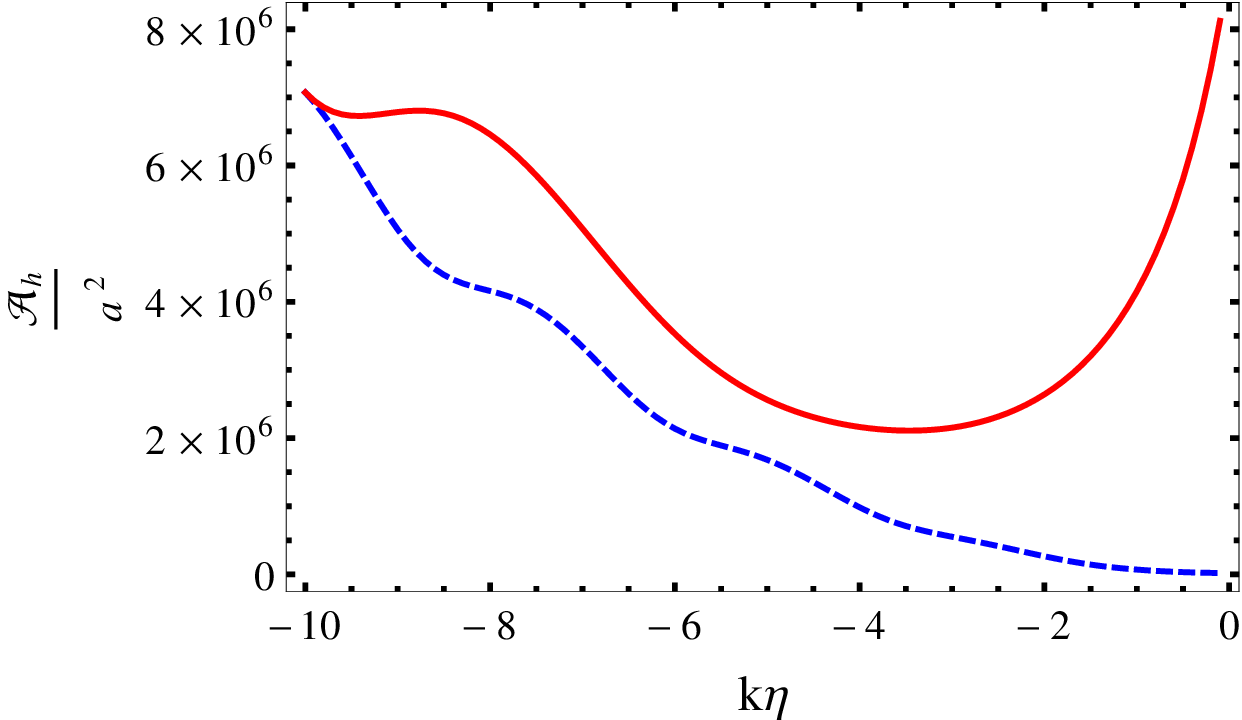}
}
{
\includegraphics[width=3.1in,height=2.1in,angle=0]{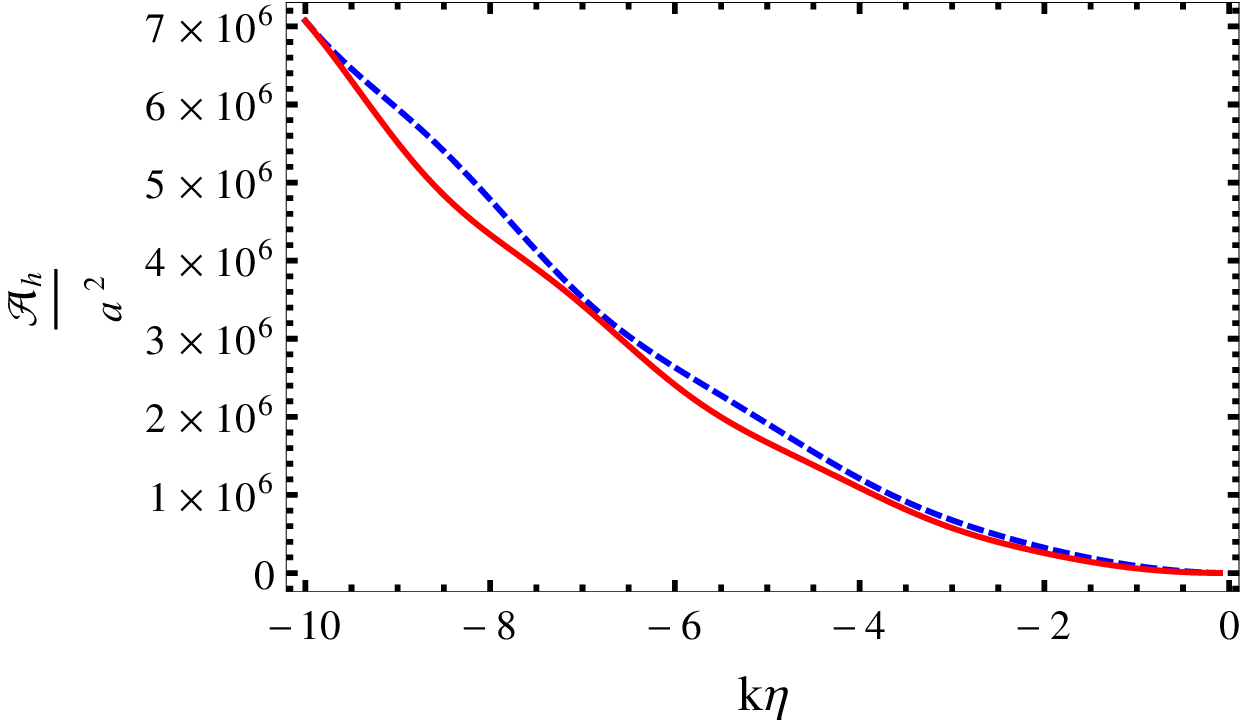}
} 
\caption{For $k=10^{-2}$ and $D=4$ The behavior of $\mathcal{A}_{h}(k\eta)/a^{2}$ has been has been shown in the plots.
The top panel shows the evolution for  $\beta D/2\alpha=2$, whereas, 
bottom panel shows the plot for $\beta D/2\alpha=-0.516$. The red solid curve 
represents positive helicity mode whereas, the blue dashed curve is for negative helicity mode.}
\label{FIG1}
\end{figure}

In the top panel of Fig.~(\ref{FIG1}), we have shown the solution for a mode $k=1/100$, $\beta D/2 \alpha=2$, which corresponds 
to $m_{1}=2$. We see that, as the modes evolve, the positive helicity mode becomes much larger than the negative helicity mode. 
Thus the dominant contribution to the energy density when $k\eta<<1$ comes from positive helicity 
modes. This is also seen for example from Eq.~(\ref{cm}) and Eq.~(\ref{Ahi}) that the ratio of power spectrum between 
$h=+1$ to $h=-1$ is $e^{2\pi n}$ and for $n=2$ this is $e^{4\pi}\simeq 3\times10^{5}$.
Thus the generated magnetic field will be significantly helical with positive helicity.

\begin{figure}[!htbp]
\centering
{
\includegraphics[width=3.1in,height=2.3in,angle=0]{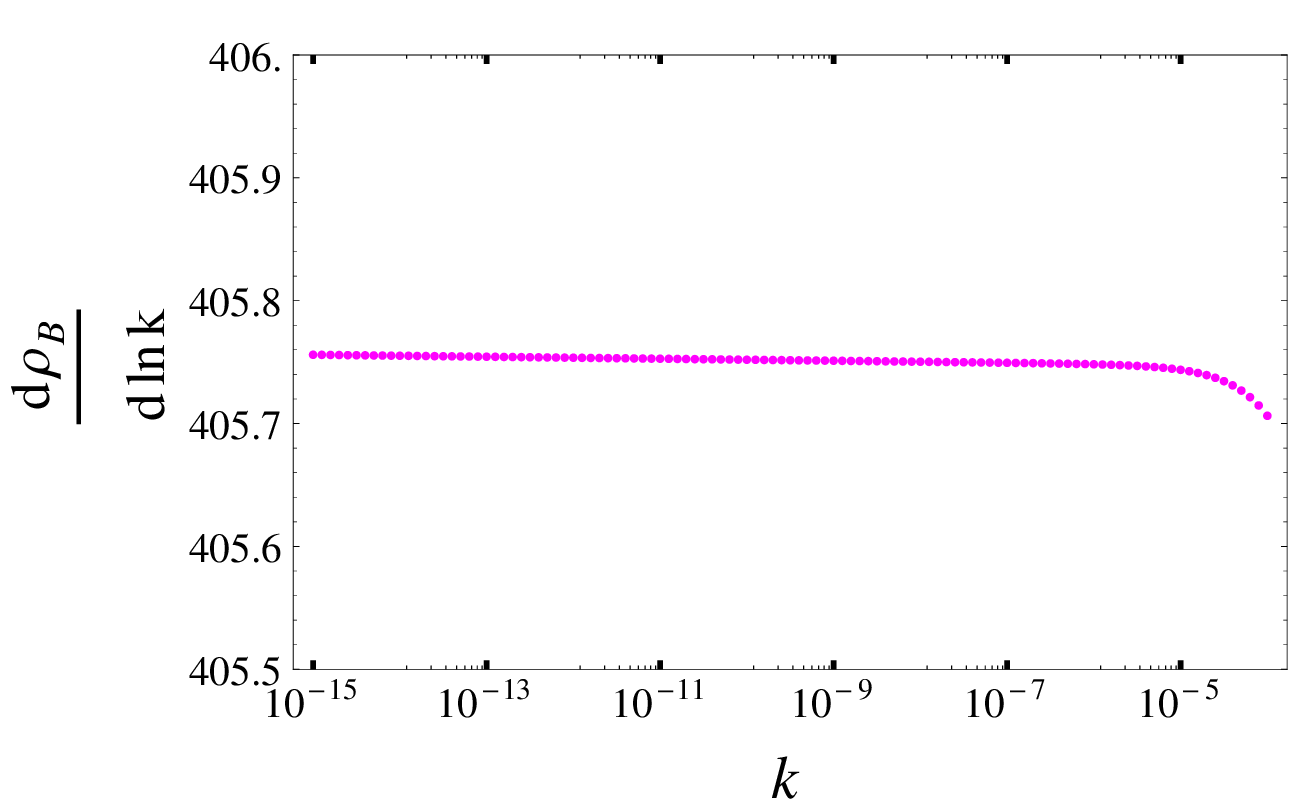}
}
{
\includegraphics[width=3.1in,height=2.3in,angle=0]{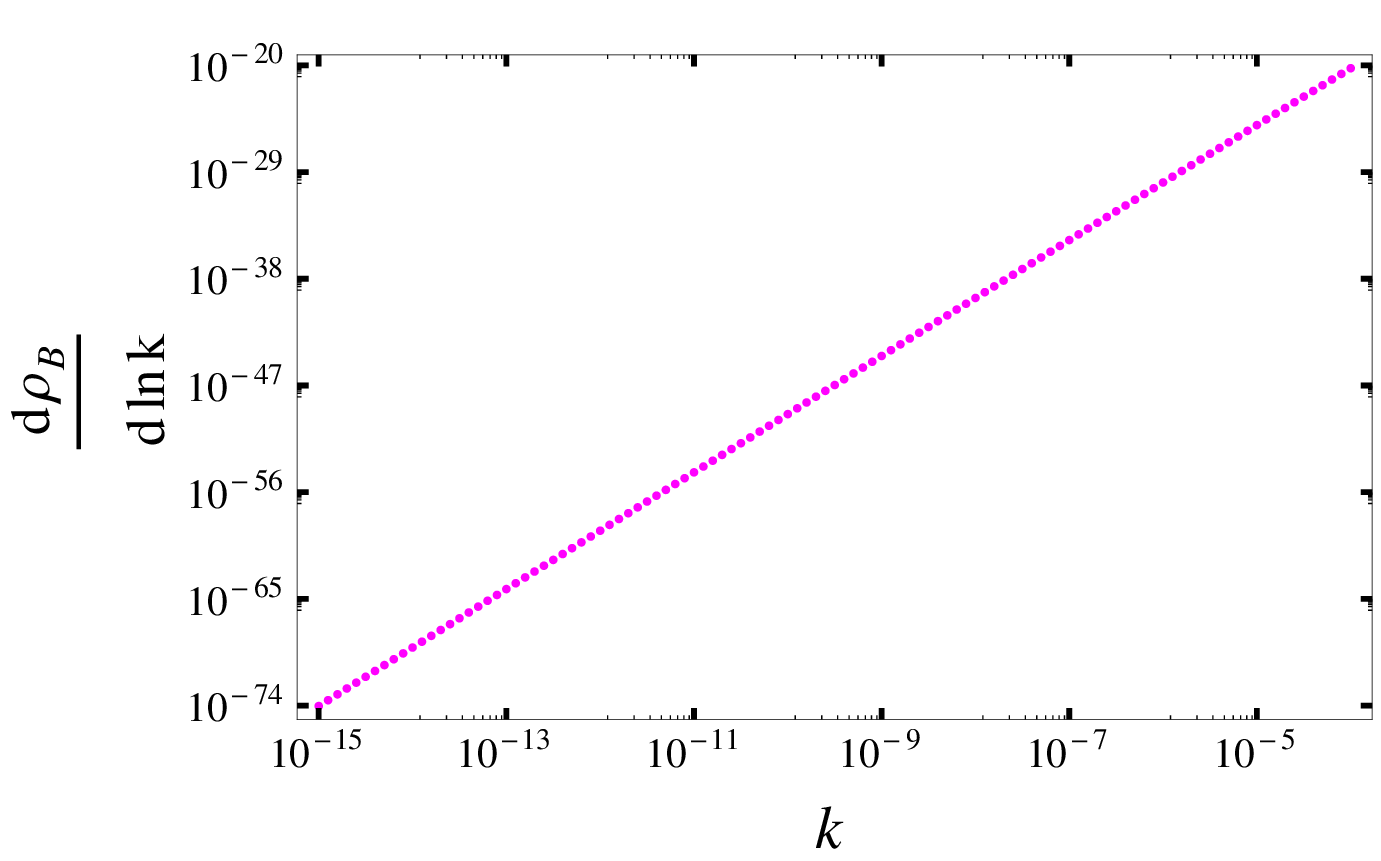}
}
\caption{The behavior of magnetic field power spectrum for $D=4$ and $m_{1}=2$ (i.e. $\beta D/2 \alpha=2$) is shown in the top panel. 
The spectrum shows scale invariance for large variation in $k$. Also for $D=4$ power spectrum for magnetic fields is 
shown for $\beta D/2\alpha=-0.516$, [bottom panel]. This is one of the cases when normal dimensions inflates while extra dimensions goes 
through contraction. The spectrum is no longer scale invariant and gives a blue spectrum with $n_{B} \simeq 4.8$.}
\label{FIG2}
\end{figure}

In the top panel of Fig~(\ref{FIG2}), we show the corresponding power spectrum for $m_{1}=2$, in the units 
of $H^4$ (which remains constant during inflation).
We have considered a range of modes corresponding to $k=10^{-15}$ to $k=10^{-4}$. The power spectrum is calculated at conformal time 
$\eta=-0.5$ for which all such modes  have already crossed outside the horizon scale. We see from this figure that one obtains a nearly 
scale invariant power spectrum for a vast range of $k$. This confirms the result from analytical solution 
that taking $m_{1}=2$ will give a scale invariant spectrum. 
The condition $\beta D/2\alpha>0$ refers to the cases for which $\alpha$ and $\beta$ have the same sign. This implies that both normal dimensions 
as well as extra 
dimensions either inflate or contract simultaneously. Such solutions were shown to exist when a nonzero $\bar{\Lambda}$ is present \cite{akis14}.

On the other hand, several other solutions with $\bar{\Lambda}=0$ were also obtained in Refs.~\cite{akis14,pahwa08}. 
In these solutions scale factors of normal and extra dimensions have opposite behavior, that is when one expands the other contracts 
leading to $\beta D/2\alpha<0$. One such solution with $D=4$ has
$\beta D/2 \alpha=-0.516$. The results for this case are shown in the bottom panel of Fig.~(\ref{FIG1}). 
The corresponding power spectrum for the magnetic field is shown in the bottom panel in Fig~(\ref{FIG2}). 
We see that the magnetic field now has a blue spectrum. A numerical fit to the power spectrum gives a spectral index $n_{B} \simeq 4.8$ for 
the same range of modes. The field strength however can be seen to be
negligible in this case compare to the one obtained for scale invariant scenario.
We note in passing that the case with $\beta D/2\alpha=-3$, also corresponds to $m_{2}=2$, and therefore gives a scale invariant power
spectrum for magnetic fields, now with predominantly negative helicity. However, such cases are not viable as they lead to unacceptably 
large electric fields \cite{my08,kandu10}. 


\section{MAGNETIC FIELD INTENSITY}\label{MF}
We now calculate the intensity of generated magnetic field at the current epoch. We consider the scale invariant case
($D=4$ and $m_{1}=2$).
From Eq.~(\ref{spectrum}) we can write,
\begin{eqnarray}
\frac{d\rho_{B}}{dlnk} &=&  \frac{1}{(2 \pi)^2}H^{4}\left|\frac{1}{\sqrt{2}}\frac{\Gamma(2m_{1}+1)}{(2m_{1}+1)!C_{m_{1}}(nh)} \right|^2 \nonumber \\
& & \simeq  2.6\times 10^{2} H^4.
\label{rhof}
 \end{eqnarray}
This is in reasonably good agreement with the estimate of $\simeq 4\times 10^{2} H^{4}$ which we obtain by directly integrating Eq.~(\ref{Ah})
numerically;  see Fig.(\ref{FIG2}). Note that in the absence of helicity the amplitude of the magnetic field spectrum is $(9/4\pi^2) H^4$ \cite{akis14}). 
Therefore in the presence of helicity the amplitude of the spectrum becomes larger by a factor of $\simeq 10^3$. 
Note the factor $(b/b_{0})^{D/2}$ is similar to a time dependent coupling functions $f(\phi)$ or $I(\eta)$ in \cite{my08,kandu10,cap14}.
However, the coupling function appears naturally in our work. 
As this factor settles to unity we recover the standard cosmology. The role of extra dimensions is important only till the end 
of inflation as extra dimensions are assumed to be frozen afterwards. Therefore, the post inflationary era the magnetic field energy density 
evolves as,
\begin{equation}
 \rho_{B}(0)=\rho_{B}(f)\left(\frac{a_f}{a_0}\right)^4.
 \label{bf}
\end{equation}
Here, $a_{f}$ and $a_0$ are the scale factors and $\rho_{B}(f)$ and $\rho_{B}(0)$ are the energy densities at the end of inflation and present 
epoch, respectively. From Eq.~(\ref{rhof}), the magnetic field intensity depends on the scale of inflation. Combining Eqs.~(\ref{rhof}) and (\ref{bf}),
we estimate that helical magnetic fields with nearly scale invariant spectrum of strength
$10^{-9}$ $G$ can be generated for $H \simeq 10^{-3}$ $M_{pl}$ (see also Ref.~\cite{akis14} for more details of the numerical estimation of the field strength). 
Further for lower scales of inflation fields weaker than $10^{-9}$ $G$ will be generated. 
The upper limit on primordial magnetic field strength, from their effects on CMB temperature anisotropy is $\simeq 3.4$ $nG$ on scale of 1 Mpc \cite{planck13}. 
From the constraints on CMB non-gaussianity, the strength of primordial magnetic field is limited to sub $nG$ level \cite{psk14,psk12,psk10,cap09}.
The lower limit is set by $\gamma$-ray observations is of order $10^{-16}$ $G$ \cite{kus11}. 
Therefore the magnetic fields that can be generated by our model are within the permissible range. 
\section{DISCUSSIONS AND CONCLUSIONS}\label{conc}
The presence of coherent magnetic fields at large scales (Mpc), even in the void regions of the IGM, indicates that these fields could have a primordial origin. 
One possibility is that they are generated during the inflationary era. However, as the background geometry is conformally 
flat, conformal invariance of the electromagnetic action action needs to be broken in order to generate a significant magnetic field.
In our earlier work \cite{akis14}, we had investigated such a possibility in the context of higher-dimensional theories. In the current work we have extended
this consideration to the possibility of generating magnetic fields which are also almost fully helical. 
Our higher-dimensional action includes the Gauss-Bonnet term which also allows one to obtain inflationary solutions without the introduction 
of scalar fields \cite{pahwa08}.

In order to study the generation of helical magnetic fields we have added a parity breaking term to the higher-dimensional electromagnetic action. 
Considering a suitable field configuration and gauge choice \cite{gio2000}, we performed a dimensional reduction of higher-dimensional electromagnetic action 
to obtain $1+3$-dimensional action. This gives rise to a dynamical coupling term as a function of scale factors of higher dimensions.
The evolution of the extra-dimensional scale factor naturally provides the requisite condition for breaking the conformal invariance of 
electromagnetic action, essential for the generation of significant magnetic fields. 

The evolution of the helical modes of vector potential in $1+3$-dimensions is described by Eq.~(\ref{Ah}).
Analytical solutions in terms of the Coulomb's functions can be obtained in special cases when $m_{i}$ in the evolution Eq.~(\ref{Ah}) is a positive 
integer. For other cases in general, one requires numerical integration of Eq.~(\ref{Ah}).
We have shown that it is possible to generate not only fully helical fields, but also one that has a scale invariant spectrum. 
Such a situation is obtained when $\beta D/2\alpha=2$ or $-3$ and corresponds to $m_{i}=2$ in Eq.~(\ref{Ah}). 
For $\beta D/2\alpha=2$, both analytical and numerical solutions show that the positive helicity modes dominate over the negative helicity modes for the 
scales which exit the horizon during inflation. The case $\beta D/2\alpha=-3$ is ruled out as it leads to unacceptably large 
electric fields. A set of solutions of the higher-dimensional Einstein's equations including $\bar{\Lambda}$ can be obtained with $D=4$, having 
$\beta=\alpha$ and thus giving $\beta D/2\alpha=2$. 
We have shown that helical magnetic fields of the order of $10^{-9}$ $G$ can then be generated in our model for $H\simeq10^{-3}$ $M_{pl}$. 
Weaker magnetic fields can be generated for further low scale inflationary models. The strength of magnetic fields generated 
by this mechanism is consistent with the constraints from CMB non-gaussianity and $\gamma$-ray observations. Note that in all the higher-dimensional models
one requires also a mechanism to freeze the evolution of the extra-dimensional scale factor. This issue needs to be investigated separately.

Recently, during the course of this work,  Refs. \cite{cap14,cheng14} have also discussed the generation of helical fields; where 
a parity violating term to the $1+3$-dimensional electromagnetic action is added with time dependent couplings. Our work differs by being set in the 
context of higher dimensional theories. In our models the conformal invariance is broken naturally by the coupling to the evolving higher-dimensional 
scale factor. 
Our analytical results match with that of Ref. \cite{cap14} wherever comparison can be made. In addition, we have included the numerical treatment 
of more general cases. Further Ref.~\cite{cap14} have limited themselves to low scales of inflation (in order to avoid strong coupling problem \cite{mukh09}) 
resulting in weaker fields with blue spectrum. 
This needs to undergo inverse cascade to explain large scale fields. 
The strong coupling problem (in most of the models) arises because a large variation of 
the coupling function (which breaks conformal invariance) is required to produce strong magnetic fields. However, in our model 
the problem of strong coupling could possibly be circumvented,  as the coupling term (which depends on the extra-dimensional scale factor) appears as 
an overall multiplicative factor to the full electromagnetic action which includes its interaction with matter. 
We hope to address this issue in more detail in the future. 


\acknowledgments{KA and TRS acknowledge the facilities at the IUCAA, Pune, for hospitality and resources to complete this work in addition to 
the facilities provided at IUCAA Resource center. KA acknowledges the UGC, India for assistance under grant AA/139/F-42/2009-10. TRS acknowledges 
CSIR, India for assisitance under grant O3(1887)/11/EMR-II}
\bibliography{helical}

\end{document}